# Are Self-explaining and Coached Problem Solving More Effective When Done by Pairs of Students Than Alone?

**Robert G.M. Hausmann[1], Brett van de Sande[2], & Kurt VanLehn[3]**
([1]bobhaus@pitt.edu, [2]bvds@pitt.edu, [3]vanlehn@cs.pitt.edu)
Pittsburgh Science of Learning Center and the Learning Research and Development Center
University of Pittsburgh, 3939 O'Hara Street, Pittsburgh, PA 15260-5179

**Abstract**

Although cognitive science has discovered several methods for increasing the learning of complex skills, such as physics problem solving, detailed examination of verbal protocols suggests there is still room for improvement. Basically, students do not always apply the meta-cognitive strategies that the instruction invites. For instance, when prompted to self-explain, students may still choose to not explain. We conjecture that most students know which meta-cognitive strategies are good and bad. When they work in pairs, they are more likely to choose the good strategies. We hypothesize that social accountability improves meta-cognitive strategy choice, which thereby improves learning. Our experiment compared individuals and pairs learning from state-of-the-art instruction. The results suggest that the dyads solved more problems and requested fewer hints during problem solving than individuals. We also discovered a new form of self-explanation, wherein students generate explanations to account for the differences between their solutions and the instructor's.

## Introduction

As any teacher knows, no matter how cleverly designed the instructional activities may be, students often devise a way to participate without learning. We will first illustrate this truism by discussing how four different methods of instruction, each with an illustrious track record, can be subverted by students. We then propose a combination of the four that could reduce the frequency of subversion and thus increase learning. This suggests an experiment which we have run using college physics problem solving as the task domain. The results agree with our predictions. After reporting the experiment, we try to explain the results in a larger context of cognitive science.

### Four successful types of instruction

When teaching problem solving, instruction often begins by presenting worked-out examples. In physics, an example consists of multiple steps leading up to the answer of a problem, where a step could be drawing a vector, defining coordinate axes, defining a variable, or writing an equation. One way to increase student learning as they study an example is to prompt them after each step to explain why the step is true, what is its role in solving the problem, how it relates to what they know already, etc. This is called *prompting for self-explanation*, and many studies have shown that it increases learning (e.g., Atkinson, Renkl, & Merrill, 2003; Chi, DeLeeuw, Chiu, & LaVancher, 1994; Taylor, O'Reilly, Sinclair, & McNamara, 2006). Of course, students can subvert such instruction by simply producing a shallow self-explanation, such as a paraphrase (Hausmann & Chi, 2002).

Another way to get students to learn from examples is to alternate them with similar problems (Trafton & Reiser, 1993). Students are told that after they have studied the example, it will be removed and they must solve a nearly identical problem. Such *example-problem alternations* have been widely shown to increase learning compared to either all-problem or all-example instruction (e.g., Atkinson, Derry, Renkl & Wortham, 2000). However, example-problem alternation can also be subverted by students. They can self-explain the example superficially then solve the problem poorly. Or they can attempt to memorize the surface features of the example and map them onto the problem (e.g., VanLehn, 1998).

*Step-based tutoring* involves a human or computer tutor that allows the student to attempt each step in solving a problem, gives feedback on each step, and gives a hint when asked a particular step (VanLehn, 2006). Many intelligent tutoring systems implement this kind of instruction, including our Andes system, which will be discussed later. Multiple studies testify to their success compared to classroom instruction (e.g., Anderson, Corbett, Koedinger, & Pelletier, 1995; VanLehn et al., 2005). However, students can subvert step-based tutoring by misusing the feedback and hints, a behavior often known as *gaming the system* (Aleven, Stahl, Schworm, Fischer, & Wallace, 2003). They either ask for so many hints that the system is constantly giving away the correct steps, or they refuse to ask for hints and rapidly guess at steps until feedback tells them that they generated a correct one (Baker, Corbett, Koedinger, & Wagner, 2004).

*Peer collaboration*, in the context of learning to solve physics problems, would involve a pair of students working together to study an example or to solve a problem. Although collaboration elicits more learning than individual work (Johnson & Johnson, 1992; Slavin, 1990), there are several ways that collaboration may fail to yield learning gains. One reason collaboration may fail is in the case where one student does most of the work while the other does little, which can be due to social loafing, domination, or both (O'Donnell & Dansereau, 1992). Another could be called collaborative floundering, wherein the students are on task and even collaborating, but are making little progress (Barron, 2003).

### Combining collaboration with three types of scaffolding

Although all four types of instruction listed above are successful, none are perfect. One hypothesis is that students may not know how to use the instruction effectively; thus, they might benefit from meta-cognitive instruction on how to learn. For instance, one study used a tutoring system to teach students how to effectively use the feedback and hints available from a step-based tutoring system (Roll et al., 2006). They found that, while the help-seeking tutor was guiding students' help-seeking behavior, they complied and learned more domain knowledge; however, as soon as the help-seeking tutor was turned off, they reverted to their old gaming behaviors. Not surprisingly, their learning returned to its earlier levels. Perhaps it is not too much of a generalization to suggest that for all four forms of instruction, most students know what the "good" meta-cognitive methods for learning are, but they sometimes choose not to apply them.

If so, then making meta-cognitive strategy choices public may embarrass students who choose a "bad" meta-cognitive strategy. To avoid embarrassment, they may choose good meta-cognitive strategies more frequently and thus learn more effectively.

This suggests having students use the best individualized instruction available—prompting for self-explanation, example-problem alternation, and step-based tutoring—but use them collaboratively, in pairs. If the collaboration was asymmetric, in that one student does most of the work and the other watches, then the student doing the work is less like to choose an embarrassing, bad meta-cognitive strategy because the other student is watching. On the other hand, if the collaboration is symmetric, with both students contributing equally, then they would have to agree on the necessity of the bad meta-cognitive strategy before using it. For instance, both students would have to agree to click rapidly on the help button, a form of gaming behavior often seen when individuals use step-based tutoring systems (Baker et al., 2004).

Thus, we hypothesize that students using these high-powered forms of scaffolding (prompting for self-explanation, example-problem alternation, and step-based tutoring) will learn more when they use them in pairs than individually. An experiment was run to test this hypothesis.

## Method

### Participants

Thirty-nine undergraduates were randomly assigned to one of two experimental conditions: self-explanation (individuals; $n$ = 11) or joint-explanation (dyads; $n$ = 14). Volunteers were recruited from several sections of a second semester physics course, which covered Electricity and Magnetism. They were recruited during the third week of the semester, with the intention that the experimental materials would coincide with their introduction in the actual physics course. The participants were paid $10 per hour. To ensure that the participants' motivation remained high during the entire two-hour session, they were offered an incentive of an additional $10 for doing well on the tests. All of the students received the bonus.

### Materials

The materials developed for this experiment were adapted from an earlier experiment. The domain selected for this experiment was electro-dynamics with an emphasis on the concept of the definition of an electric field, which is expressed as a vector equation: $\boldsymbol{F} = q\boldsymbol{E}$. This particular topic is an important concept for students to master because it represents their first exposure to the idea that a field can exert a force on a body.

To instruct the participants, several materials were developed. Four electrodynamics problems were created, which are representative of typical problems found at the end of a traditional physics textbook chapter. The problems covered a variety of topics, including the definition of the electric field; Newton's first and second law, the weight law, and several kinematics equations. Each of the four problems was implemented in Andes[1] (VanLehn et al., 2005).

Andes is a step-based tutoring system, in the sense that it offers feedback and help for each problem-solving step. Users can ask Andes for hints on an incorrect entry or for hints on the next problem-solving step. The first, or top-level hint, is very general and abstract. The purpose is to remind the student what action to take. If students ask for a second hint, it is more specific and direct, and the goal is to teach the student the justification for taking a step. Finally, the last hint, or the *bottom-out hint*, tells the student explicitly what action to take (see Fig. 1 for an example).

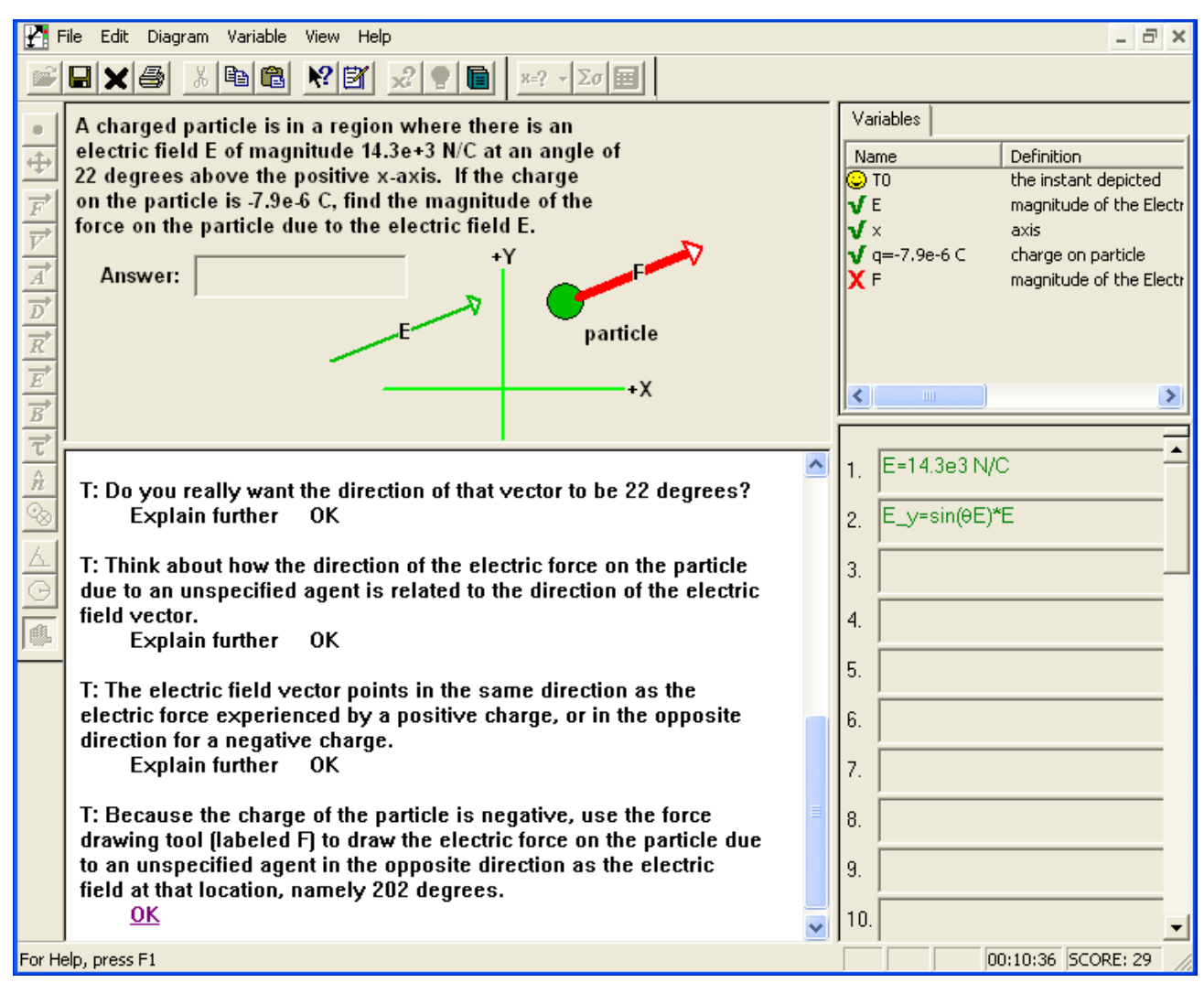

Figure 1. A screen shot of Andes showing a hint sequence. The final "bottom-out" hint gives explicit instructions.

[1] http://www.andes.pitt.edu/

Andes was chosen as the learning context for two reasons. First, its design allowed for both coached problem solving and the presentation of video-based examples. Second, because Andes is a step-based tutor, it has the ancillary benefit of focusing conversations on the step itself. This may help increase the productivity of collaboration because it helps students avoid collaborative floundering.

The first problem served as a warm-up problem because none of the students had prior experience with the Andes interface. The problems grew in complexity in that additional principles were needed to solve successive problems. Dyads solved all of the problems as a pair.

In addition to the problems, three examples were created in collaboration with two physics instructors at the U.S. Naval Academy. The examples contained a voice-over narration of an expert solution to the problems, and they were structured such that they were isomorphic to the immediately preceding problem.

## Procedure

The procedure consisted of several activities. All were done either individually or in pairs, depending on the condition. First, participants read a short description of a self- or joint-explanation. Then they watched a short, introductory video on the Andes user interface. Afterwards, they used Andes to solve a warm-up problem. The experimenter was available to answer any user-interface questions; however, he was not allowed to give away any domain-specific information. During problem solving, the students had access to Andes' flag feedback (correct/incorrect), hint sequences, and equation cheat sheet.

Once the students submitted a final answer, they then watched and explained an example of an expert solution of an isomorphic problem. The example solutions were broken down into steps, and at the conclusion of each step the students were prompted to explain it aloud; all participants wore headset microphones to record their explanations. When done with their explanations, the participants clicked a button to go onto the next step. Only the cover story and given values differed between the problem-solving and example problems.

Note that the order of solving and studying examples differs from traditional research on example-studying (Atkinson, Derry, Renkl, & Wortham, 2000). In the present experiment, students attempted to solve a problem *first*, and then studied an isomorphic example *second*. The students alternated between solving problems and studying examples until all four problems were solved and all three examples were studied, or until two hours elapsed.

## Measures

Several dependent measures were used to assess problem-solving performance. We used the log files generated by Andes to count the number of errors the students made while solving problems. In particular, we were interested in the number of hints, and especially the number of bottom-out hints requested. We will present each measure, as well as an excerpt from the dialogs in which the dyads were studying an example.

# Results

Andes assists student while solving problems so that all of the problems can be solved correctly, albeit with more or less guidance. Because everyone who finishes a problem gets the right answer, process measures, such as latencies, errors, and hint requests, must be used to assess competence.

The experiment was capped at two hours, which means the students were asked to stop even if they had not finished all the problems. Therefore, counts of errors, hints, and duration are reported as rates, where we divided each measure by the number of correct entries. This normalization takes into account different problem solving techniques, as well as the possibility of students not finishing all of the problems.

## Correct-entry latencies

Working in a group can sometimes result in *process loss* (Steiner, 1972), where groups are slower than individuals because they need extra time to coordinate and synthesize their ideas. To test if process loss occurred, we measured the latency as the number of seconds between correct entries, which is an indication of problem-solving efficiency. The dyads ($M$ = 80.24, $SD$ = 16.90) demonstrated a shorter average latency than the individuals ($M$ = 101.44, $SD$ = 19.80). This difference was statistically reliable, with a large effect size, $F(1, 23) = 8.35$, $p = .008$, $d = 1.21$. This result suggests that the dyads did not suffer process loss; instead, they made faster progress on entering the correct steps than the individuals.

Although many training experiments require students to finish a fixed set of problems and use time-to-completion as a dependent measure, our experiment was capped at two hours. Thus, the number of problems and number of correct steps completed were used as dependent measures of competence. The dyads finished more problems than the individuals (see Table 1). Moreover, the dyads ($M$ = 49.21, $SD$ = 4.61) also entered more correct entries than the individuals ($M$ = 44.36, $SD$ = 6.70). The difference was both statistically and practically significant, $F(1, 23) = 4.60$, $p$ = .04, $d$ = .90.

Table 1: The number of students who finished each problem.

| | **Individuals** | **Dyads** | $\chi^2$ (1, $N$ = 25) |
|---|---|---|---|
| Warm-up | 11/11 | 14/14 | . |
| Problem 1 | 11/11 | 14/14 | . |
| Problem 2 | 9/11 | 14/14 | $\chi^2 = 2.77$, $p = .10$ |
| Problem 3 | 3/11 | 10/14 | $\chi^2 = 4.81$, $p = .03$ |

## Error rate

The number of errors a student makes when solving a problem is influenced by how well they learned during earlier problem solving and example studying. An error, defined in the context of solving problems with the Andes system, was any student entry that turned red (incorrect).

The entries may be marked as incorrect by Andes under three conditions. The first is when the entry is ill-formed. For instance, units must be included on dimensional quantities. Secondly, an error is flagged if the entry is untrue of the physical situation (e.g., a vector is drawn in the wrong direction). The third condition is when Andes does not recognize the entry as a step along a particular solution path. Problems typically have multiple solution paths, and Andes knows virtually all of them. However, students occasionally make a true entry that is not needed for any solution. In that case, Andes also flags the entry as incorrect.

For the present analysis, we defined the error rate as the ratio of incorrect to correct entries. Therefore, larger numbers indicate that more incorrect entries were made before entering the correct step.

The dyads ($M$ = .71, $SD$ = .31) demonstrated a lower error rate than the individuals ($M$ = .97, $SD$ = .49). However, there was only a trend for this relationship, with a medium effect size, $F(1, 23) = 2.61$, $p = .12$, $d = .68$. This suggests that the dyads were only slightly less likely to enter an incorrect entry than the individuals.

## Hint and bottom-out hint requests

Although the error rates were not statistically reliable, there is a way, in an intelligent tutoring system, to avoid making errors. Instead of attempting a step and failing, students can take preventative action by asking for a hint. Because the hints were written in a graded fashion, each correct step can have multiple hints (usually three).

Consistent with the error-rate results, the individuals ($M$ = 2.26, $SD$ = 1.52) requested more overall hints from the Andes tutoring system than the dyads ($M$ = .99, $SD$ = .82). The differential use of the help system was statistically reliable, with a large effect size (see the left side of Fig. 2), $F(1, 23) = 7.20$, $p = .01$, $d = 1.13$.

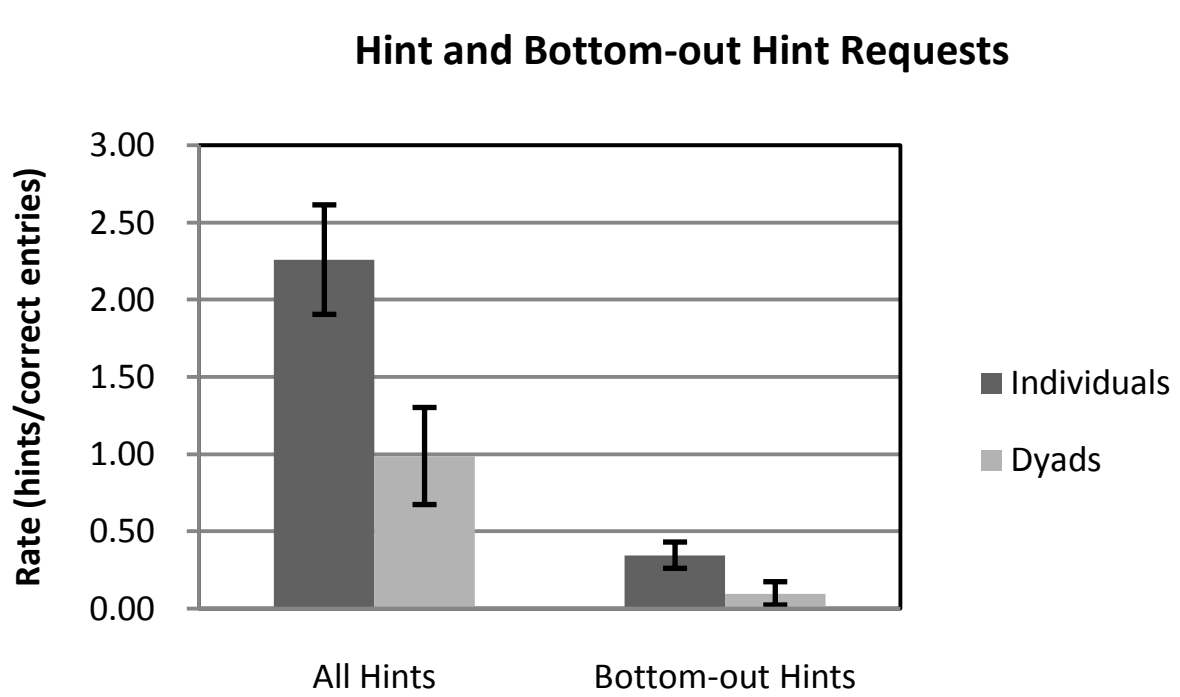

Figure 2. Average number of hints and bottom-out hints requested during problem solving.

Moreover, if students are completely stuck on a step, then they can ask for a “bottom-out hint,” which will tell them exactly what to enter. The individuals ($M$ = .34, $SD$ = .40) requested more bottom-out hints per correct entry than the dyads ($M$ = .10, $SD$ = .14). This difference was statistically reliable, with a large effect size (see the right side of Fig. 2), $F(1, 23) = 4.71$, $p = .04$, $d = .91$.

Both the overall and bottom-out hint usage results suggest that the dyads were able to fix their incorrect entries and impasses with less assistance from the Andes help system than the individuals.

## Analyses of example-studying behavior

Our hypothesis is that working in pairs changes the mixture of meta-cognitive strategies during both problem solving and example studying. The preceding section focused on problem solving, where it appears that dyads are asking for fewer bottom-out hints and working faster than individuals, as predicted. This section summarizes our progress on analyzing the example-studying behavior.

Because transcription is not yet complete, we can only give informal observations. The first is that the pairs produced much more talk than the solos. This is consistent with our prediction that pairs should produce more and deeper self-explanations than individuals, but there are many other possible explanations as well.

Although we have conducted many studies of self-explanation in physics, we were surprised to see a new type. When students were prompted to self-explain an example’s solution, they often compared it to the solution that they produced on the immediately preceding problem.

To illustrate how this interaction unfolds, consider the following excerpt (see Table 2). The dialog takes place after the dyad has solved the warm-up problem. During their solution, the pair was required to write an equation that describes the magnitude of an electrical force on a charged particle found in an electric field (i.e., $\boldsymbol{F} = q\boldsymbol{E}$). In Andes, this equation can be written in two different ways. It can be expressed in terms of its components (i.e., $F_x = q*E_x$ & $F_y = q*E_y$) or in terms of its magnitude (i.e., F = abs(q)*E). The dyad in this example chose to express the definition of the electric field in terms of its components. Because the sought quantity for the warm-up problem asks for the magnitude, an additional step is required to find the resultant force (i.e., the Pythagorean Theorem). The video-based example presented the step as using the magnitude-only equation.

The interaction begins with Amy recognizing that the method presented in the video is another valid way of taking the same step that they used during their problem solving (line 2), and her partner agrees (line 3). But the conversation does not end with the recognition that there are two ways to complete the same step. Amy also produces an explanation for *why* the magnitude-only equation also works (line 4). Erin takes the explanation one step further by observing that it will save an additional problem-solving step by avoiding the Pythagorean Theorem to find the resultant vector (line 7). On their next problem-solving attempt to write this formula (not shown in the Table 2), they used the absolute value. However, they mixed up the order of the electric field and force (i.e., E = abs(q)*F). This suggests that the example, and the dyad’s ensuing dialog, had an impact on their problem-solving behavior.

Table 2: An example of a dyad collaboratively comparing their previous solution to an expert solution.

| Line | Speaker | Statement |
|---|---|---|
| 1 | Video | "Now that the direction of the electric force has been indicated, we can work on finding the magnitude. We write this equation, in the equation window" [ Video shows: Fe=abs(q)*E ]. |
| 2 | Amy | That works, too. |
| 3 | Erin | Yeah. |
| 4 | Amy | I guess it's absolute value because the direction is already taken care of, though. |
| 5 | Erin | Yeah. |
| 6 | Amy | We're just looking for magnitude anyway, but |
| 7 | Erin | Yeah, that-, I guess that sorta gets rid of the whole square-root step that we did. |

This kind of self-explanation has not yet been mentioned in the literature, but in retrospect it seems to be a natural consequence of our instructional procedure. As mentioned earlier, students are presented with pairs of isomorphic problems. They solve the first member of the pair with the aid of Andes, and then they study an instructor's solution of the second member of the pair. As they study the example, they often compare "his" solution to their own (the voice presenting the video-based example was male).

This new kind of self-explanation could be beneficial. For instance, if students had trouble with a particular step during problem solving, then they may be more likely to self-explain the step deeply when studying the example; this may remedy their confusion. On the other hand, if their step agrees with the example's step, then they may think that they already know everything there is to know about this step, and thus not self-explain it as deeply (a problem-solver's version of the *illusion of knowing*, Glenberg, Wilkinson, & Epstein, 1982). This new kind of self-explanation needs to be studied further.

## Discussion

Cognitive science has discovered several methods of instruction that are more effective than baseline methods (e.g., reading or unassisted problem solving). These effective methods include prompting for self-explanation, example-problem alternation, and step-based tutoring. Although all have produced large effect sizes, detailed examination of verbal protocols and computer-tutor log files suggest room for improvement. For instance, even when prompted to self-explain, students do not always do so.

We discovered that having students do these activities in pairs seems to have increased learning compared to doing the activities as individuals in that dyads took less time per correct step and asked for less help per correct step. Given that the three instructional methods were already known to be significantly more effective than baseline methods of instruction, the fact that we are able to show any improvement at all is rather surprising.

However, it must also be said that the experimental procedure, and in particular, the self-imposed cap of two hours, limits our claims. The students alternated between solving problems with a tutoring system (Andes) and studying examples with prompting. We used their performance on the tutoring system to measure competence. Compared to individuals, the dyads were able to solve more problems, take less time per correct step, and request less assistance from Andes. There was also a trend for dyads to make fewer errors.

However, we did not use conventional pre- and post-testing, which would have required students to solve problems individually, without the aid of a tutoring system. We could not fit such testing into the 2-hour period and still have time for instruction. Thus, this experiment needs to be replicated with traditional assessments in order to check that learning gains, as measured conventionally, did in fact differ between conditions.

Our main reason for running an unconventional design was to test for the hypothesized changes in meta-cognitive strategies. Compared to individuals, we expected the pairs to produce more and deeper self-explanations during example-studying and to use the help system in more appropriate ways during tutored problem solving. We have indirect evidence that the hypothesized differences occurred during problem solving, in that pairs asked for bottom-out hints much less frequently than individuals. However, we do not yet have evidence for the hypothesized differences in self-explanation, other than the fact that pairs produced more words than individuals, and that difference could have multiple explanations. Analyses of the transcripts and log files are continuing.

We did discover an unanticipated behavior in both the pairs and individuals. Example-problem pair instruction usually gives students two isomorphic problems: the first one is an example and the second one requires the student to produce a solution similar to the one shown in the example. When students work problems without any help, this order is essential, because otherwise they could have great difficulty solving the problems. In earlier studies, we observed a great deal of unproductive copying behavior when examples are presented before problems (VanLehn, 1998), so we tried problem-example pairs instead of example-problem pairs. That is, the first member of each pair was a problem to be solved by the student, and the second was a similar problem with its solution. This was feasible because students solved the problems with the aid of a tutoring system. Although we did not anticipate it, this dramatically changed the students' behavior during example studying. Their self-explanations often compared the example's solution to their own problem solving. Because this feature of our procedure was not manipulated, we do not know whether it helped or hurt learning. Comparing example-problem pairs to problem-example pairs would be an interesting follow-up experiment.

In conclusion, we have found signs of increased effectiveness from a combination of a partially-successful instructional method (peer collaboration) with three successful instructional methods (prompting for self-explanation, example-problem alternation, and step-based tutoring; all usually done by students working alone). We found preliminary evidence of a difference in the mix of meta-cognitive strategies, consistent with our hypothesis that working in pairs tends to decrease the frequency of poor meta-cognitive strategies. To put it succinctly, it appears that social accountability increased the frequency of effective learning strategies, which in turn increased learning gains.

## Acknowledgments

This work was supported by the Pittsburgh Science of Learning Center, which is funded by the National Science Foundation award number SBE-0354420. The authors are deeply indebted to Robert Shelby and Donald Treacy for their assistance in developing the instructional materials.